\documentclass[preprint2]{aastex}
\usepackage{amsfonts}
\usepackage{amssymb}
\usepackage{subfigure}
\usepackage{graphicx,color}

\usepackage{ulem}
\usepackage{amsmath}

\newcommand{\psec}{\ensuremath{\, {\rm s}^{-1}}}

\newcommand{\km}{\ensuremath{\,{\rm km}}}

\newcommand{\Mpc}{\ensuremath{\,{\rm Mpc}}}

\newcommand{\MHz}{\ensuremath{\, {\rm MHz}}}

\begin{document}

\title{{\LARGE The Global 21 cm Absorption from Cosmic Dawn with Inhomogeneous Gas Distribution}}

\author{{\large Yidong Xu\altaffilmark{1}, Bin Yue\altaffilmark{1}, Xuelei Chen\altaffilmark{1,2,3}\\}}
\altaffiltext{1}{Key Laboratory for Computational Astrophysics, National Astronomical Observatories, 
Chinese Academy of Sciences, Beijing 100101, China}
\altaffiltext{2}{School of Astronomy and Space Science, University of Chinese Academy of Sciences, Beijing 100049, China}
\altaffiltext{3}{Center for High Energy Physics, Peking University, Beijing 100871, China} 

%\maketitle
\begin{abstract}
We make an analytical estimate of the maximum 21 cm absorption signal from the 
cosmic dawn, taking into account the inhomogeneity of gas distribution in the
intergalactic medium (IGM) due to non-linear structure formation. 
%While the over-dense gas surrounding the halos
% could make more contribution to the absorption, this gas is also adiabatically heated to a higher
%temperature, and as a result the maximum global 21 cm absorption depth is reduced by about 40\%
The gas located near halos is over-dense but adiabatically heated, 
while the gas far from halos is under-dense and hence cooler. The cumulative effect of 
adiabatic heating and cooling from this gas inhomogeneity results in a reduction in the 
maximum global 21 cm absorption depth by about 40\%
as compared with homogeneous IGM %intergalactic medium 
model, assuming 
saturated coupling between the spin temperature 
of neutral hydrogen (HI) and the adiabatic gas kinetic temperature.
\end{abstract}


  %--- Methods: analytical}

\maketitle

\section{Introduction}\label{Intro}

The EDGES experiment (Experiment to Detect the Global Epoch of reionization Signature;
\citealt{2010Natur.468..796B}) has recently announced the discovery of an absorption 
feature in the global spectrum of the sky around 78 MHz \citep{2018Natur.555...67B}. 
If this is due to the global 21 cm absorption signal from the 
cosmic dawn \citep{1997ApJ...475..429M,2004ApJ...602....1C,2008ApJ...684...18C,2006MNRAS.371..867F}, 
it amounts to 21 cm brightness temperature
$\delta T_{\rm 21} = -500^{+200}_{-500}\, {\rm mK}$ (99\% confidence level, \citealt{2018Natur.555...67B}), 
which is much deeper than what is 
typically expected from standard models of the diffuse intergalactic medium (IGM), indeed even deeper than the largest
possible absorption by the adiabatically cooled primordial gas. To explain this 
excess absorption,  various models are developed, e.g. to cool the gas by interaction with cold 
dark matter (e.g. \citealt{2018Natur.555...71B}; \citealt{2018arXiv180210094M}; \citealt{2018arXiv180210577F,
2018arXiv180303091B,2018arXiv180309734S,2018arXiv180310671H,
2018arXiv180401092M}),  or to increase the background radiation intensity (e.g. \citealt{2018arXiv180207432F,
2018arXiv180301815E,2018arXiv180303245F}).

Previous calculation of the expected global signal level considered only the diffuse IGM
with the mean cosmic density, or with the large-scale density field from relatively low-resolution simulations
(e.g. \citealt{2017MNRAS.472.1915C}). However, the matter density surrounding halos is enhanced 
\citep{2004MNRAS.347...59B,2006ApJ...645.1001P,2006ApJ...649..579B},
which could make a significant contribution to the total absorption signature. On the other hand, 
as the virial temperatures of halos at the redshifts of interest are typically higher than the CMB temperature,
the neutral gas inside halos could produce weak 21 cm emission signals which compensate part of the absorption 
signal \citep{2009MNRAS.398.2122Y}.
Here we re-evaluate the total 21cm global signal by focusing on the cold gas surrounding the halos which is
neither virialized nor heated by astrophysical or exotic radiations. 

While the absorption depth is proportional to the gas density, adiabatic heating 
would increase the temperature and leads to only a small increase in the absorption from over-dense regions
 in the IGM \citep{2018Natur.555...71B}.
The gas in the voids on the other hand, would produce a less weak absorption due to the opposite effects.
The overall effect of the inhomogeneous gas distribution is not yet quantified.
Using a high redshift point source as the background, \citet{2011MNRAS.410.2025X} computed the 21 cm absorption
signals from individual minihalos and dwarf galaxies, including both the gas contents inside and around 
the halos. With similar algorithm for computing the optical depth, 
but using the CMB as the background radio source, we can derive the global 21 cm signal from
the cumulative absorption from the inhomogeneous gas distribution within and around halos 
arising from non-linear structure formation.
We find that before the occurrence of significant X-ray heating in the early Universe, 
the maximum global 21 cm absorption depth is reduced 
when compared with the case of homogeneous IGM,
thanks to  the adiabatic heating effect of the over-dense gas surrounding the halos.

Throughout this paper, we assume the $\Lambda$CDM model and adopt the 
Planck 2015 cosmological parameters \citep{2016A&A...594A..13P}:
 $\Omega_b = 0.0485$, $\Omega_c = 0.259$, $\Omega_\Lambda = 0.692$, 
$H_{\rm 0} = 67.81\km\psec \Mpc^{-1}$, $\sigma_{\rm 8} = 0.8149$ and 
$n_{\rm s} = 0.9677$. However, the effect discussed here are not sensitive to the
parameter values.

\section{The 21 cm absorption line from the gas around one halo}\label{one_halo_line}

We begin by calculating the 21 cm absorption line profile from the neutral hydrogen
 surrounding an {\it isolated} halo.
Note that we also include the absorption from the gas inside halos assuming
collisional ionization equilibrium \citep{2011MNRAS.410.2025X}, but
this part of gas only has a small contribution to the total absorption.
As we shall find from the calculations below, the main contribution to the optical depth is from the gas outside of halos
with the adiabatic temperature much lower than the viral temperature.
The 21 cm optical depth of an isolated object can be derived from the integral of the
absorption coefficient along the line of sight
\citep{2002ApJ...579....1F,2011MNRAS.410.2025X}:
\begin{align}
\tau (\nu) &= \frac{3\,h_{\rm P}\,c^3 A_{10}}{32\, \pi^{3/2}
k_{\rm B}}\, \frac{1}{\nu_{10}^2} \nonumber  \\
&\times \int_{-\infty}^{+\infty} {\rm d}x\, \frac{n_{\rm HI}(r)} {b(r)T_{\rm
S}(r)}\, \exp\left[\,-\, \frac{(u(\nu)-\bar
v(r))^2}{b^2(r)}\,\right] ,
\label{Eq.tau}
\end{align}
where $h_{\rm P}$ is the Planck constant, $c$ is the speed of light,
$A_{10} = 2.85 \times 10^{-15} \psec$ is the Einstein coefficient
for the spontaneous decay of the 21 cm transition, and 
$k_{\rm B}$ is the Boltzmann constant.
In the integrant, $n_{\rm HI}$, $T_{\rm S}$, and $b$ are the neutral hydrogen
number density, the spin temperature, and the Doppler parameter of the gas,
respectively. $b(r) = \sqrt{\,2\,k_{\rm B}T_{\rm K}(r)/m_{\rm H}}$, $u(\nu) \equiv c\,
(\nu-\nu_{10})/\nu_{10}$ with $\nu_{10} =
1420.4 \MHz$ being the rest frame frequency of the 21 cm transition, 
and $\bar v(r)$ is bulk velocity of gas
projected to the line of sight at the radius $r$. The coordinate $x$ is related to the radius $r$ by $r^2 =
(\alpha\, r_{\rm vir})^2 + x^2$, where $\alpha$ is the impact
parameter of the penetrating line of sight in units of the halo virial radius $r_{\rm vir}$.

The cold over-dense gas around the halos could 
absorb the CMB photons and enhance the 21 cm absorption with respect to that due to only the diffuse IGM.
However, even in the absence of X-ray heating from the first astrophysical sources, the over-dense gas surrounding
halos is adiabatically heated according to $T_{\rm K} \propto \rho^{2/3}$, and due to dearth of metals, the cooling is very 
slow. This counteracts and reduces the density enhancement effect.
To estimate the maximum level of absorption, here 
 we assume saturated coupling between the spin temperature
and the gas kinetic temperature by the Ly$\alpha$ radiation via the Wouthuysen-Field effect 
(\citealt{1952AJ.....57R..31W,1958PIRE...46..240F}; \citealt{2006MNRAS.367..259H}), i.e. $T_{\rm S} = T_{\rm K}$.

In addition to the enhanced density, the gravitational potential of a halo also induces
infall of the surrounding gas \citep{2004MNRAS.347...59B,2006ApJ...645.1001P,2006ApJ...649..579B}, and
the gas outside the virial radius has a velocity which is determined by the competition 
between the infall and the Hubble flow.
The gas density and infall velocity profiles are calculated with the spherical collapse ``Infall Model'' \citep{2004MNRAS.347...59B}
\footnote{The ``Infall Model'' code is publicly available at
http://wise-obs.tau.ac.il/$\tilde{\;\:}$barkana/codes.html.}.

\begin{figure}[htb]     
\centering{
\includegraphics[scale=0.38]{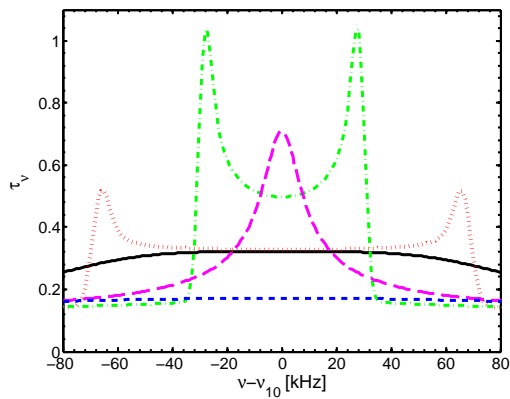}
\caption{The optical depth profiles for a few line of sights passing through the gas around 
an isolated halo with mass $M = 10^8 M_\odot$ at $z=17$, assuming adiabatic gas temperature.
 The impact parameters of the different curves are $\alpha$ = 0 (solid black), 1 (dotted red), 3 (dot-dashed green), 
 10 (long-dashed magenta), and $\alpha$ = 30 (short-dashed blue), respectively.}
\label{Fig.profile}
}
\end{figure}

The optical depth profiles of a halo with mass $10^8 M_\odot$ at redshift $z=17$ are plotted in Fig.~\ref{Fig.profile}
for a few impact parameter values. The peak optical depth is significantly enhanced for lines of sight
penetrating through the gas just outside the virial radius (the dotted, dot-dashed, and long-dashed lines for 
$\alpha = 1$, 3, and 10, respectively), as compared to the optical depth
caused by the IGM of the mean density (the short-dashed line for $\alpha = 30$).

\begin{figure}[htb]
\centering{
\includegraphics[scale=0.38]{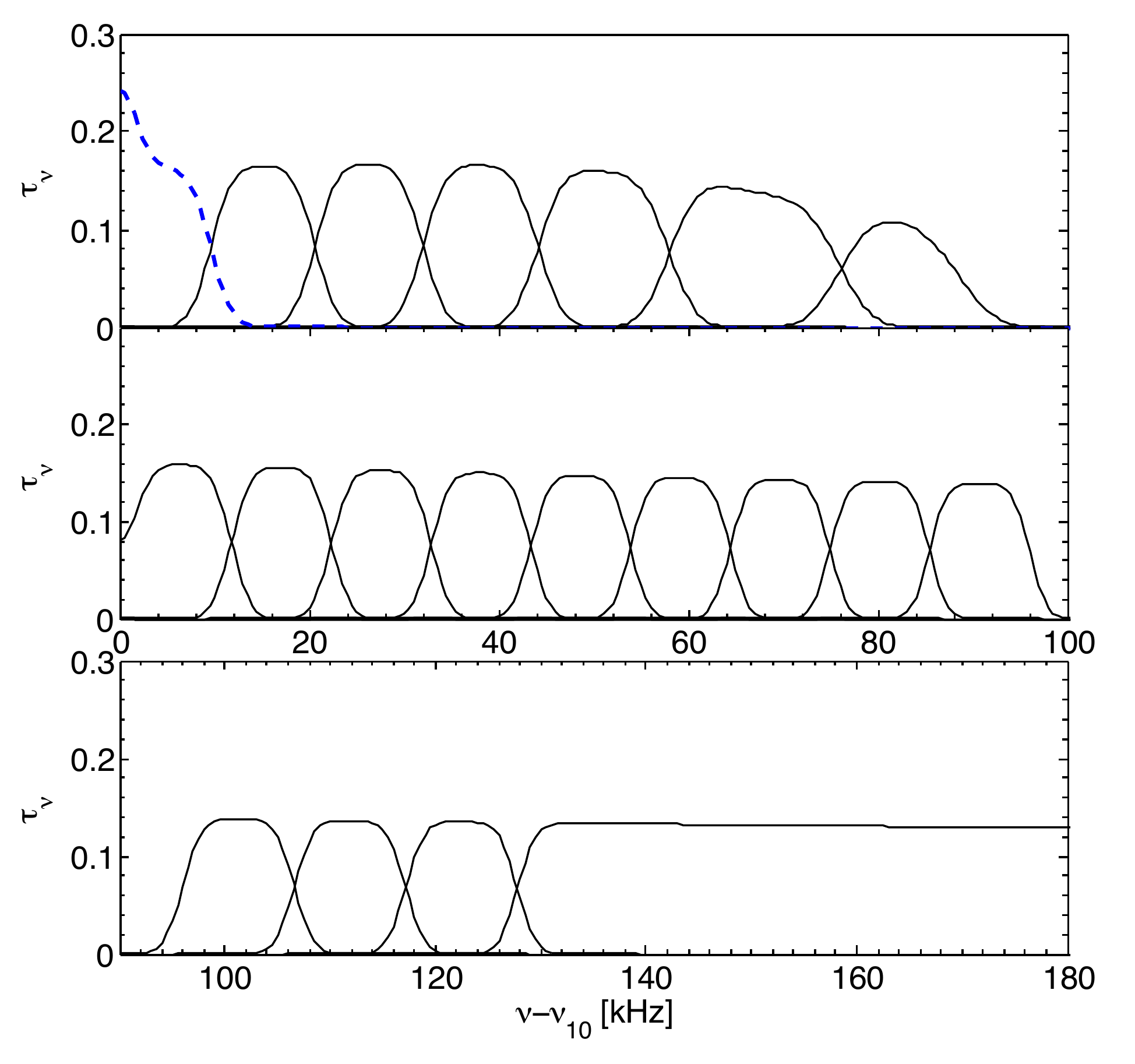}
\caption{Contributions to the optical depth from the gas at different radii,
for a line of sight penetrating through the center of an isolated halo of mass $10^8 M_\odot$ at redshift 17. 
{\it Upper panel:} the solid lines from right to left correspond to the absorptions by segments of 
(1 -- 2) $r_{\rm vir}$, (2 -- 3) $r_{\rm vir}$, (3 -- 4) $r_{\rm vir}$, (4 -- 5) $r_{\rm vir}$, (5 -- 6) $r_{\rm vir}$ 
and (6 -- 7) $r_{\rm vir}$, respectively, and the dashed line represents the absorption by the segment of (7 -- 8) $r_{\rm vir}$. 
{\it Central panel:} the 9 solid lines from left to right correspond to the absorptions by segments of 1 $r_{\rm vir}$ each 
starting from 8 $r_{\rm vir}$.
{\it Bottom panel:} the 3 solid lines from left to right correspond to the absorptions by segments of 1 $r_{\rm vir}$ each 
starting from 17 $r_{\rm vir}$, and the last curve represents the integral absorption from 20 to 100 $r_{\rm vir}$.}
\label{Fig.segments}
}
\end{figure}

The line profiles in Fig.\ref{Fig.profile} can be understood by
noting the different contributions from the gas located at different
radii \citep{2011MNRAS.410.2025X}. 
 Let $\nu_p(r)$ be the peak frequency of
optical depth generated by gas located at radius $r$, 
the absorption is shifted from the line center by
\begin{equation}
\nu_p(r) - \nu_{\rm 10} \,=\, \frac{\bar v(r)}{c}\, \nu_{\rm 10}.
\end{equation}
Because of the infall, the absorption line is significantly broadened, dramatically
exceeding the thermal broadening.
The segmental contributions to the optical depth from the gas located at different radii
are plotted in Fig.\ref{Fig.segments}. 
The gas located just outside the $r_{\rm vir}$ has the
largest infall velocity, and the corresponding absorption lies at the
largest distance to the line center in the upper panel. Its optical depth
is however a bit suppressed because of the higher temperature of the denser gas. 
As the radius increases, the $\tau_\nu$ profile gets closer to the line
center because of the decreasing infall velocity. 
At the turn-around point, where the gas velocity changes from infall-dominated
to Hubble-flow-dominated, the two
absorption lines created by the two segments on both sides of the
halo converge into one at the line center, and they contribute
substantially to the central optical depth. After that, the $\tau_\nu$
profiles that come from larger radii leave the line center again,
and when it goes out of the region influenced by
the halo's gravity, and the density drops to the cosmic mean
value, we recover the IGM mean optical depth.

Therefore, for a line of sight penetrating through an isolated halo at cosmic
dawn with a small impact parameter, the optical depth is significantly enhanced, 
not only because of the enhanced density, but also because of
the infalling velocity which re-distributed the absorption depth.
In order to correctly determine the optical depth and the profile of the 21 absorption lines
from individual halo surroundings, 
it is important to take into account both the effect of enhanced density and the infalling 
velocity of the gas.
However, when considering the overall absorption by the gas surrounding all halos, the 
infalling velocity which acts as a re-distributor, may not have such a significant effect. 
The adiabatic heating/cooling of the over-dense/under-dense gas, on the other hand,
 may have a cumulative effect.
Note that here we have assumed that the gas follows the dark matter distribution according to 
the infall model, and the gas fraction equals to the cosmic mean baryonic fraction
everywhere. In the following, we will investigate the relative effects of enhanced density, the 
infalling velocity, as well as the adiabatic heating/cooling, on the cumulative absorption 
from the gas around non-isolated halos, with the total gas content normalized.

\section{The 21 cm spectrum from cumulative gas around all halos}\label{spectrum}

The radiative transfer equation in the Rayleigh-Jeans limit gives the emergent 21 cm
brightness temperature:
\begin{align}\label{Eq.Tb}
T_{\rm b} &= T^{\rm out}_{\rm CMB} + T^{\rm out}_{\rm em} \nonumber  \\
&= T_{\rm CMB} \, e^{-\tau_{\rm tot}} + \int^{\tau_{\rm tot}}_0 T_{\rm S}(\tau)\, e^{-\tau}\, {\rm d}\tau,
\end{align}
where $T^{\rm out}_{\rm CMB}$ is the contribution from the background radiation, 
and $T^{\rm out}_{\rm em}$ is the contribution from the gas emission.
Here $T_{\rm CMB}$ is the CMB temperature at redshift $z$, and
$\tau_{\rm tot}$ is the integrated optical depth from all gas in the Universe
that could contribute to the 21cm absorption at the corresponding frequency.
We have to account for the gas surrounding all halos, from the regions with enhanced
density near the halos to the regions with lower density faraway.
Accordingly, the gas is adiabatically heated or cooled depending on the local density.
In the case of homogeneous spin temperature of HI, Eq.(\ref{Eq.Tb}) reduces to
\begin{equation}
T_{\rm b} \,=\, T_{\rm CMB} \, e^{-\tau_{\rm tot}} + T_{\rm S} \, (1 \,-\, e^{-\tau_{\rm tot}}).
\end{equation}
However, here we have to use the original form of Eq.(\ref{Eq.Tb}) in order to 
account for the inhomogeneity in the gas temperature.

In the above we find that the infall velocity is important for broadening 
the absorption line produced by each halo. 
As a consequence, the gas with the bulk velocity of $\bar v(r)$ can contribute to the 
absorption at the central 21 cm frequency at $z$, which is at a comoving distance 
$\Delta l =(1+z)\bar{v}(r)/H(r)$ away from it, so for the 21cm optical depth corresponding to a certain redshift $z$,
one needs to integrate the contribution up to distance $\Delta l_{\rm max} $ away, which
is determined by the maximum infall velocity which depends on the halo mass.
The integrated optical depth of the 21 cm absorption at an intrinsic frequency $\nu_{10}$ 
at redshift $z_c$ is then
\begin{align}
\tau_{\rm tot} (z_c) &= \frac{3\,h_{\rm P}\,c^3 A_{10}}{32\, \pi^{3/2}
k_{\rm B}}\, \frac{1}{\nu_{10}^2} \nonumber  \\
&\times \int_{-\infty}^{+\infty} {\rm d}l (z)\; \int_{M_{\rm min}}^{M_{\rm max}} 
\frac{{\rm d}n}{{\rm d}M}(M,z) {\rm d}M \nonumber  \\
&\times \int_{0}^{\alpha_{\rm max}} 2\,\pi\, r^2_{\rm vir}(M,z) \,(1+z)^2\, \alpha {\rm d}\alpha \nonumber  \\
&\times \int_{-x_{\rm max}}^{+x_{\rm max}}  {\rm d}x\; \frac{n_{\rm HI}(r)} {b(r)T_{\rm
S}(r)}\, \exp\left[\,-\, \frac{(u(\nu)-\bar v(r))^2}{b^2(r)}\,\right] .
\label{Eq.integ_tau}
\end{align}
Here $\nu$ is related to $\nu_{10}$ by $\nu/(1+z) = \nu_{10}/(1+z_c)$, 
$x_{\rm max} = r_{\rm vir} \sqrt{\alpha_{\rm max}^2 - \alpha^2}$,
and ${\rm d}n\,/\,{\rm d}M$ is the halo mass function, for which we adopt the Sheth-Tormen form
 \citep{2002MNRAS.329...61S}.
We integrate the absorption by the gas content out to a maximum impact parameter 
$\alpha_{\rm max}$, which is set to be half of the mean halo distance determined by the minimum
halo mass under consideration; 
the gas located at farther than this distance would probably be
influenced by the gravitational potential of a neighboring halo, and this part of gas is
taken into account as contribution from the neighboring halo surrounding.

For a patch of HI gas from one halo, the emission term in the emergent brightness
can be written as
\begin{align}
T^{\rm out}_{\rm em,1h} 
&= \int_{-x_{\rm max}}^{+x_{\rm max}} T_{\rm S}(x)\, e^{-\tau(x)}\, \kappa_\nu(x)\, {\rm d}x \nonumber  \\
&= \frac{3\,h_{\rm P}\,c^3 A_{10}}{32\, \pi^{3/2}
k_{\rm B}}\, \frac{1}{\nu_{10}^2} \nonumber  \\
&\times \int_{-x_{\rm max}}^{+x_{\rm max}} \,e^{-\tau(x)}\, 
\frac{n_{\rm HI}(r)} {b(r)}\, \exp\left[\,-\, \frac{(u(\nu)-\bar v(r))^2}{b^2(r)}\,
\right]  {\rm d}x,
\label{Eq.Tex_1h}
\end{align}
where
\begin{align}
\tau(x) &= \frac{3\,h_{\rm P}\,c^3 A_{10}}{32\, \pi^{3/2}
k_{\rm B}}\, \frac{1}{\nu_{10}^2} \nonumber  \\
&\times \int_{-x_{\rm max}}^{x} \,
\frac{n_{\rm HI}(r')} {b(r')T_{\rm S}(r')}\, \exp\left[\,-\, \frac{(u(\nu)-\bar v(r'))^2}{b^2(r')}\,
\right]  {\rm d}x'.
\label{Eq.tau_p}
\end{align}
Adding up the emission from all halos,
the total emission term in Eq.(\ref{Eq.Tb}) is then
\begin{align}
T^{\rm out}_{\rm em} 
&= \frac{3\,h_{\rm P}\,c^3 A_{10}}{32\, \pi^{3/2}
k_{\rm B}}\, \frac{1}{\nu_{10}^2} \nonumber  \\
&\times \int_{-\infty}^{+\infty} {\rm d}l (z)\; \int_{M_{\rm min}}^{M_{\rm max}} 
\frac{{\rm d}n}{{\rm d}M}(M,z) {\rm d}M \nonumber  \\
&\times \int_{0}^{\alpha_{\rm max}} 2\,\pi\, r^2_{\rm vir}(M,z) \,(1+z)^2\, \alpha {\rm d}\alpha \nonumber  \\
&\times \int_{-x_{\rm max}}^{+x_{\rm max}} \,e^{-\tau(x)}\, 
\frac{n_{\rm HI}(r)} {b(r)}\, \exp\left[\,-\, \frac{(u(\nu)-\bar v(r))^2}{b^2(r)}\,
\right]  {\rm d}x.
\label{Eq.Tex_1h}
\end{align}
The observed 21 cm signal is the differential brightness temperature relative to
the CMB, i.e.
\begin{equation}\label{Eq.dT21}
\delta T_{\rm 21} (z_c) \,=\, \frac{T_{\rm b}(z_c) \,-\, T_{\rm CMB}(z_c)}{1\,+\,z_c}.
\end{equation}

\begin{figure}[htb]
\centering{
\includegraphics[scale=0.4]{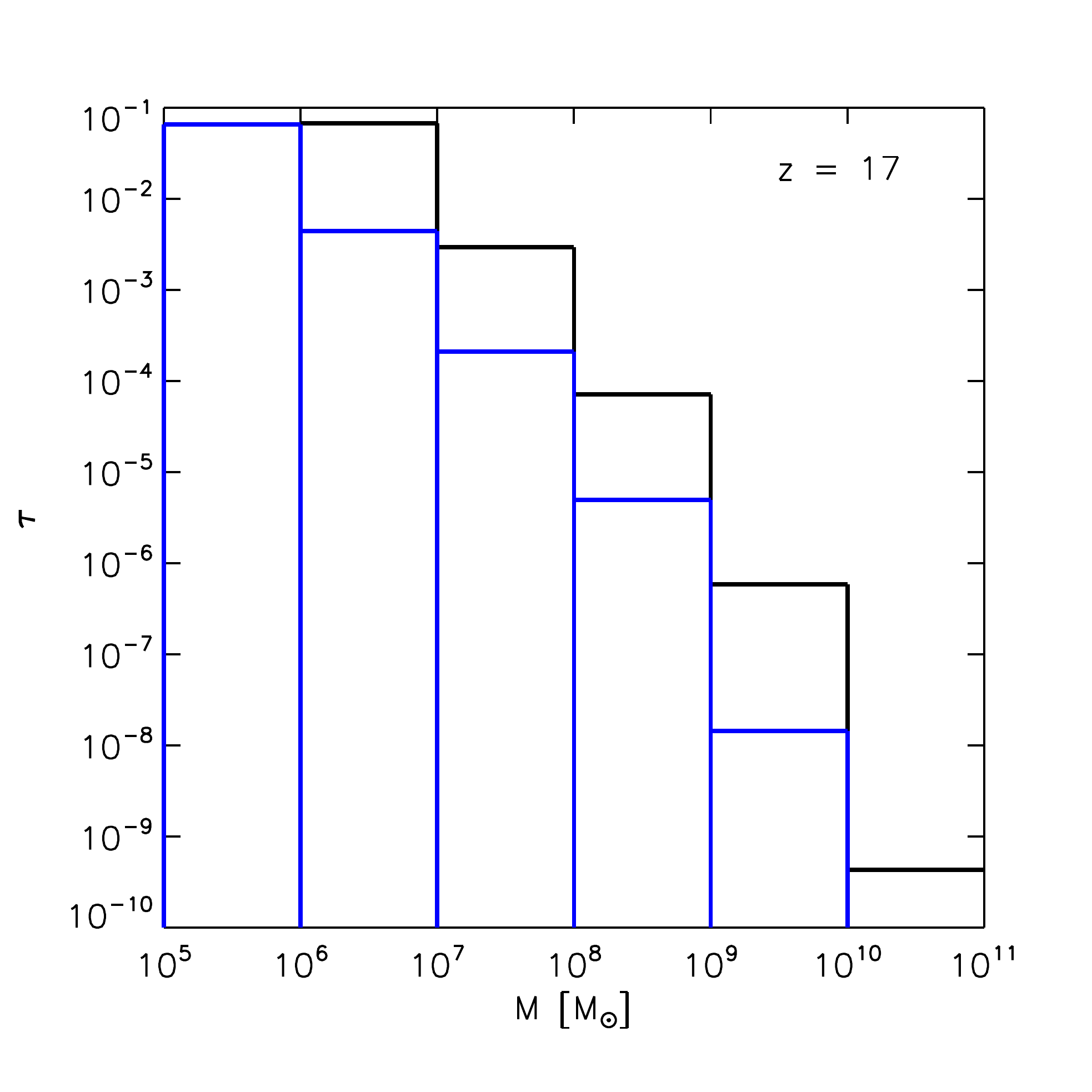}
\caption{The optical depth contributed from the gas surrounding halos of different mass ranges at redshift 17. 
The blue histogram shows the contribution of various halos assuming $M_{\rm min} = 10^5 M_\odot$, 
and the black histogram shows the result with $M_{\rm min} = 10^6 M_\odot$.}
\label{Fig.mass_contrib}
}
\end{figure}

The 21 cm signal depends on the neutral gas content of matter in and around halos.
Small halos have shallow gravitational potential and could hardly retain 
enhanced gas density around them, while
very massive halos would have high probability of hosting the first stars or galaxies at cosmic dawn 
and get the surrounding gas ionized. Beyond redshift 7, for an early X-ray background less than 
$20\%$ of the intensity today, the characteristic mass $M_{\rm C}$ at which halos on average could retain half of their
baryons is $\sim 10^6 M_\odot$  \citep{2011MNRAS.410.2025X}.
Halos with lower mass could also have gas content in and around them, but
we take this mass limit as the minimum mass for a halo that could induce {\it enhanced} gas density
{\it outside} of it. The gas around lower mass halos is taken into account by the integration up
to the maximum impact parameter, well into the IGM.
In the following, we consider halos with masses in the range
of [$10^6 M_\odot$ -- $10^{11} M_\odot$]
which cover the characteristic halo mass and most of the halos that 
could retain cold and neutral gas around them.
We will also show results with $M_{\rm min} = 10^5 M_\odot$, in order to investigate 
the effect of this uncertain mass limit.
The higher mass halos have dense infalling gas around them, and even the
adiabatic heating could raise the gas temperature in the nearest region above the 
CMB temperature at the time. 
This effect, in combination with 
the much lower number density of massive halos, results in negligible 
contribution from the gas surrounding halos more massive than 
$10^{10} M_\odot$ (as seen from Fig.~\ref{Fig.mass_contrib}), 
and we conservatively set the upper limit to be $10^{11} M_\odot$.
We also tried including the contribution from the gas in and around halos up
to $10^{12} M_\odot$ assuming no ionization, and found that 
it made a negligible correction to the total optical depth.

\begin{figure}[htb]
\centering{
\includegraphics[scale=0.4]{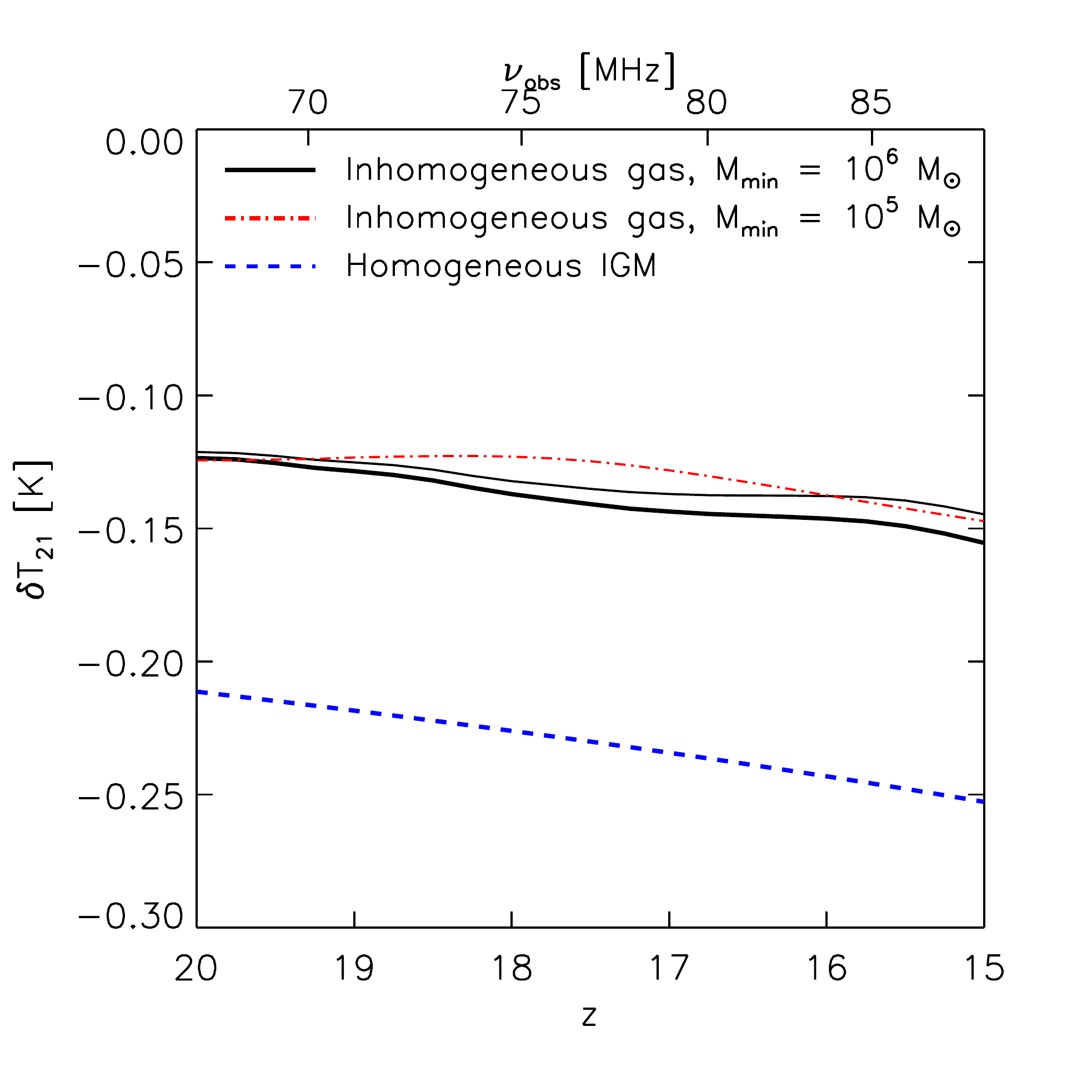}
\caption{The 21 cm global spectrum from the adiabatic gas before reionization 
assuming saturated Ly$\alpha$ coupling between the HI spin temperature and the gas kinetic 
temperature. The thick solid line shows the maximum signal from the inhomogeneous gas
due to non-linear structure formation in the default model with $M_{\rm min} = 10^6 M_\odot$,
the thin solid line shows the signal with normalized gas content, and
the dot-dashed line shows the signal with normalized gas content assuming $M_{\rm min} = 10^5 M_\odot$.
For comparison, the maximum signal from the homogeneous IGM is plotted with dashed line.}
\label{Fig.spectrum_Tk}
}
\end{figure}

Using the mean density profile predicted by the ``infall model'', 
the different contributions to the optical depth from  the gas surrounding  halos of different mass ranges
are plotted with the histogram in Fig.~\ref{Fig.mass_contrib}.
We find that the gas surrounding low mass halos dominates the absorption.
Combining the background radiation attenuated by the total optical depth $T^{\rm out}_{\rm CMB}$
and the gas emission term $T^{\rm out}_{\rm em}$, the thick solid line in 
Fig.~\ref{Fig.spectrum_Tk} shows the maximum 21 cm absorption signal, or the minimum 21 cm
brightness temperature evolution during the cosmic dawn, 
when the inhomogeneous gas distribution 
with inhomogeneous temperature is taken into account.
The previous prediction in which all gas is homogeneously distributed in the IGM,
is plotted with the thick dashed line.
We find that the maximum 21 cm global spectrum signal expected in the standard model
is reduced by a factor of about 40\% as compared with previous prediction
with homogeneous IGM, due to the inhomogeneity in the gas distribution.

Note that there is large uncertainty in the density profile around halos, even in the 
ideal case predicted by the infall model (see the large scatter in Fig. 3 of 
\citealt{2004MNRAS.347...59B}). In the above calculation, we have neglected the 
halo clustering, and the infall model predicts the density profile within a biased region 
with an initial over-density. In a more realistic situation, the halos tend to cluster in
over-dense regions, and the distances between halos in these regions are smaller than
the mean halo distance estimated from the mean halo number density. If we integrate 
the gas content up to half of the mean halo distance around each halo, we would 
over-count the total gas amount.
Here we introduce a normalization factor to the mean
density profile predicted by the model, i.e. $\rho(r) = C\, [1 + \delta(r)] \, \bar{\rho}$,
where the normalization factor $C$ is determined by requiring the conservation of 
the total gas content in the Universe.  
We find that for the mean density profile predicted by the ``Infall Model'', 
this normalization factor ranges from $\sim 0.87\, (z = 20)$ to $\sim 0.67\, (z = 15)$
if the $M_{\rm min}$ that can induce enhanced surrounding density is $10^6 M_\odot$, 
or  from $\sim 0.69 \, (z = 20)$ to $\sim 0.58 \, (z = 15)$
if $ M_{\rm min} = 10^5 M_\odot$. 
It is smaller than 1 as expected.
The results with normalized gas content are shown in Fig.~\ref{Fig.spectrum_Tk} 
with thin solid line for $M_{\rm min} = 10^6 M_\odot$ and thin dot-dashed line
for $M_{\rm min} = 10^5 M_\odot$, respectively.
The signal is further reduced due to the normalized gas content, but the decrement
is fairly small because of the roughly $\rho^{1/3}$ scaling of the optical depth
in the adiabatic case.
Note that the detailed gas density profile depends on the halo formation history and 
should vary from place to place depending on the local gravitational potential.

We also investigate the maximum signal from inhomogeneous gas density
but with homogeneous temperature
without adiabatic heating or cooling effect.
We find that the inhomogeneous distribution of the gas density alone weakly decreases
the global 21 cm absorption (by $\sim 3\%$ for $M_{\rm min} = 10^6 M_\odot$ and 
by $\sim 7\%$ for $M_{\rm min} = 10^5 M_\odot$ at redshift 17), whereas in combination with the
adiabatic heating effect reduces the absorption signal by $\sim 40\%$.
%the gas near halos is over-dense and adiabatically heated, and the gas in voids is adiabatically
%cooled, and

The uncertainty in the minimum halo mass that could induce a higher density in the surrounding 
gas does affect the contribution of the absorption from different halo surroundings as seen in 
Fig.~\ref{Fig.mass_contrib}. However, by integrating the contribution from all the HI gas in the Universe,
the overall effect of the adiabatic heating/cooling due to the inhomogeneity is not very sensitive to the
assumed $M_{\rm min}$, and the results assuming $M_{\rm min} = 10^6 M_\odot$ and that with 
$M_{\rm min} = 10^5 M_\odot$ differ by less than 10\%.
We also test the effect of the infall velocity, and find that by
integrating over the gas surrounding all halos, the infall velocity induced by individual halos shifts 
the contribution of the absorption from different halos, but only has a negligible effect on the
overall absorption depth.

\section{Conclusions and discussions}\label{conclusions}

In this work, we have re-calculated the maximum signal of the global 21 cm spectrum from the cosmic
dawn taking into account of the inhomogeneous distribution of gas in the IGM due to non-linear structure formation.
Assuming adiabatic gas temperature and saturated Ly$\alpha$ coupling between the HI spin
temperature and the gas kinetic temperature, and using the density profile of halo surroundings
predicted by the infall model \citep{2004MNRAS.347...59B}, 
we integrate the absorption from the
gas surrounding almost all halos that could retain over-dense neutral and cold gas nearby. 
We find that the maximum 
global 21 cm signal that could possibly be reached in the standard model is even weaker than
previously thought, by a factor of $\sim 40\%$, mainly because of the adiabatic heating of the 
 inhomogeneous gas around non-linear structures. 
 This result further escalates the tension between the standard model
prediction and the observational evidence of the deep absorption signal \citep{2018Natur.555...67B}.

One caveat in the above analytical estimation is that the infall model we adopted was developed
for isolated halos. In fact, neighboring halos could modify the density distribution and the velocity profile
around the halos due to the tidal effect that is inevitable in the realistic environments, 
and hence affect the predicted optical depth.
In addition, halo clustering, for which we have not accounted, could further reduce the accuracy 
of the model prediction. The gas fractions around halos of different masses are also quite uncertain, depending on
the local gravitational potential and the photon-evaporation effect once the first stars formed.
More accurate gas distribution in the Universe requires high-resolution hydrodynamic simulation.
Therefore, the result presented here should only be taken as qualitative.
However, the reduced 21 cm absorption is caused by the inhomogeneous distribution of the gas and the
related adiabatic heating effect, which is inevitable no matter what density profile we adopt.
A more precise estimation of the evolution of the global 21 cm signal also requires detailed 
modeling of the first star formation and the evolution in the Ly$\alpha$ background.
We plan to implement more realistic calculation of the early Ly$\alpha$ background level
and the HI spin temperature in a future work.

Recently, \citet{2018arXiv180402406V} have found that in the absence of the first ionizing sources, 
there is extra heating by the CMB, 
which may induce a $\sim 10\%$ correction on the gas temperature, and will further reduce the signal.
We have not included this effect in this work, but combining these two effects, we conclude that
the maximum global 21 cm signal from the cosmic dawn should be weaker than the previous estimate. 

Any X-ray heating in the early Universe would be an additional effect to further decrease the 
absorption level. Depending on the nature of the X-ray sources (stellar sources, X-ray binaries, 
QSOs, or other exotic sources), the evolutional behavior of its effect on the global 21 cm signal 
would be different from the adiabatic heating effect that is expected to be smoother in redshift. 
Also, the angular dependence of the 21 cm fluctuations may be used to distinguish the X-ray 
heating effect from the adiabatic heating effect. The adiabatic heating effect is correlated with 
the spatial distribution of predominantly small halos, while the temperature fluctuations from 
X-ray heating would dominate the 21 cm fluctuations on larger scales depending on the 
X-ray sources \citep{2018arXiv180803287R}.

\acknowledgments
This work is supported by the Chinese Academy of Sciences (CAS) 
Strategic Priority Research Program XDA15020200, 
the CAS Frontier Science Key Project QYZDJ-SSW-SLH017, 
the National Natural Science Foundation of China (NSFC) key project grant 11633004, 
the NSFC-ISF joint research program No. 11761141012, 
and the MoST 2016YFE0100300. 
BY acknowledges the support of  the Hundred Talents (Young Talents) program from the 
CAS, the NSFC grant 11653003 and 
the NSFC-CAS  joint fund for space scientific satellites No. U1738125.

\bibliography{references}
\bibliographystyle{hapj}

\end{document}